\documentclass[aps,twocolumn,eqsecnum,showpacs]{revtex4}
\usepackage{amsthm}
\usepackage{amsmath}
\usepackage{latexsym}
\usepackage{amsfonts}
\usepackage{amssymb}
\usepackage{color}
\usepackage{bbm,dsfont}

\newtheorem{proposition}{Proposition}

\theoremstyle{definition}

\newtheorem{example}{Example}

\newcommand{\real}{\mathbb R} 
\newcommand{\complex}{\mathbb C} 
\newcommand{\half}{\frac{1}{2}} 
\renewcommand{\mod}[1]{|#1|} 

\newcommand{\hi}{\mathcal{H}} 
\newcommand{\ip}[2]{\left\langle\,#1\,|\,#2\,\right\rangle} 
\newcommand{\kb}[2]{|#1\,\rangle\langle\,#2|} 
\newcommand{\no}[1]{\left\|#1\right\|} 
\newcommand{\tr}[1]{\textrm{tr}\left[#1\right]} 
\newcommand{\id}{I} 

\newcommand{\va}{\mathbf{a}} 
\newcommand{\vb}{\mathbf{b}} 
\newcommand{\vsigma}{\boldsymbol{\sigma}} 
\newcommand{\vj}{\vec{j}} 

\newcommand{\A}{\mathcal{A}}
\newcommand{\B}{\mathcal{B}}
\newcommand{\C}{\mathcal{C}}
\newcommand{\D}{\mathcal{D}}
\newcommand{\E}{\mathcal{E}}
\newcommand{\X}{\mathcal{X}}

\newcommand{\R}{\mathcal{R}}%
\newcommand{\Rsame}{\R_{\rm same}}
\newcommand{\Rdiff}{\R_{\rm diff}}

\newcommand{\obs}{\mathcal{O}}
\newcommand{\cO}{{\mathcal{O}}}

\newcommand{\be}{\begin{eqnarray}}
\newcommand{\ee}{\end{eqnarray}}
\def\r{\rangle}
\def\l{\langle}

\begin{document}

\title[Discrimination of quantum observables]{Discrimination of quantum observables using limited resources}
\author{Mario Ziman$^{1,2}$, Teiko Heinosaari$^{1,3}$}
\affiliation{$^{1}$Research Center for Quantum Information, Slovak Academy of Sciences, D\'ubravsk\'a cesta 9, 845 11 Bratislava, Slovakia\\
$^{2}$Faculty of Informatics, Masaryk University, Botanick\'a 68a, 602 00 Brno, Czech Republic\\
$^{3}$Department of Physics, University of Turku, Finland
}
\date{\today}
\email{ziman@savba.sk}
\email{temihe@utu.fi}

\begin{abstract}
We address the problem of unambiguous discrimination and identification
among quantum observables. We set a general framework and investigate
in details the case of qubit observables. In particular, we show that perfect 
discrimination with two shots is possible only for sharp qubit observables 
(e.g. Stern-Gerlach apparatuses) associated with mutually orthogonal directions. 
We also show that for sharp qubit observables associated to nonorthogonal directions unambiguous discrimination with an inconclusive result is always possible.
\end{abstract}
\maketitle

\section{Introduction}\label{sec:intro}

Quantum theory is a statistical theory in its predictions 
and therefore, a measurement result must be understood as a probability 
distribution over possible measurement outcomes. 
It could thus seem that an individual
outcome obtained in a single experimental run can give hardly any 
information. The only conclusion one can draw from a single outcome 
is that the system under investigation is in a state which gives 
nonzero probability for the obtained outcome. 

However, there are situations in which an individual outcome may provide significant amount of  information. In such cases the investigated questions have only finite number of possible answers. On one hand we can have a very nontrivial 
apriori information or assumption about the quantum object, or its property. 
For example, in communication tasks 
we usually assume that the encoding of the information is known to a receiver. 
On the second hand, our goals may not be to identify 
the objects completely, or to quantify the properties perfectly, but rather to
investigate the validity of some hypothesis \cite{QDET76,Chefles00}. For instance, we may want to verify whether the system is in an excited state. In such cases an individual measurement outcome may provide us with sufficient arguments to make the right conclusion.

The previous issue is well recognized in the realm of state discrimination
\cite{Chefles00}.   
By measuring an informationally complete observable (i.e. performing a complete 
state tomography) one will find out the state of the system. 
However, this requires a large number of identically prepared systems as one needs to know the measurement outcome frequencies. A particular sequence of measurement outcomes does not reveal the state of the system.

On the other hand, it is well-known that two known orthogonal states 
can be discriminated even if only a single copy of the system is available.
There are also many weaker variations of quantum state discrimination problems and they can be divided into two classes: 
i) minimum error discrimination \cite{QDET76} 
(conclusions are only statistical) and 
ii) unambiguous discrimination \cite{Ivanovic87,Dieks88,Peres88} 
(conclusions are error-free but an inconclusive result is possible). 

In this work we investigate perfect and unambiguous discrimination of quantum observables when only small number of probe systems are available. The problem is, briefly, to identify an unknown observable using some suitably prepared probe systems. It is assumed that the unknown observable is one from a finite set of known observables. We show that discrimination is indeed possible if the set of known observables is of the specific type. 

Our analysis proceeds in the following way. In Section \ref{sec:observables} 
we recall some concepts which are essential in our investigation. 
In Section \ref{sec:problem} we formulate the perfect discrimination problem 
and derive some general conditions for a solution.  In particular, we
give a complete solution  for perfect discrimination of two qubit observables with two shots. Section \ref{sec:unambiguous} describes the general 
unambiguous identification problem of observables 
and a solution for sharp qubit observables is presented. Finally, in Section \ref{sec:conclusions} we summarize the obtained results.

A similar problem has been investigated by Ji et al. in 
\cite{JiFeDuYi06}, but they assumed that the unknown measurement apparatus 
has labeled outcomes (c.f. Section \ref{sec:equivalence}) and they mostly studied schemes were either known measurement or unitary operator is also performed. 
In our investigation it is assumed that only measurements with the unknown apparatus are allowed. 

\section{Observables}\label{sec:observables}
\subsection{Positive operator measures}
Loosely speaking, an observable is something which attaches a probability distribution 
of measurement outcomes to each state of the system.
Two observables are, by definition, different if they lead to 
different probability distributions of measurement outcomes at 
least in some state. In quantum mechanics observables are 
conventionally represented by normalized positive operator 
measures \cite{QTOS76},\cite{PSAQT82}. 
We briefly recall this concept.

Let $\hi$ be the Hilbert space describing the system under investigation. The states of the system are positive trace class operators with trace one. Let $\Omega=\{\omega_1,\ldots,\omega_k \}$ be the set of possible measurement outcomes (in this work we only consider observables with finitely many outcomes).  A normalized positive operator measure (POVM) on $\Omega$ is a mapping $\A:\omega_j\mapsto \A_j$ such that each $\A_j$ is a positive operator in $\hi$ and $\sum_{j=1}^k \A_j=\id$. If the system is in a state $\varrho$, then the probability of getting the outcome $\omega_j$ when measuring $\A$ is $\tr{\varrho\A_j}$.

\subsection{Identifying observables}
\label{sec:iden_obs}
Observables give raise to probability distributions, not only on
individual outcomes, but also on sequences of outcomes if the measurement
is repeated. The identification of observables is based on the identification 
of these probability distributions. For a fixed state $\varrho$, consider 
two probability distributions $p$ and $q$ corresponding to observables 
$\A$ and $\B$, respectively. Our goal
is to discriminate $\A$ and $\B$ perfectly, hence to uniquely identify the 
probabilities $p,q$ from particular experimental outcome, or sequences 
of outcomes. Each sequence of outcomes $\omega_1,\dots,\omega_n$, 
therefore, must be uniquely associated with 
one of the observables. In other words, the probability 
vanishes at least for one of the observables, say $\A$. 
If this is the case and the sequence of outcomes  
is indeed measured, then we can conclude with certainty that the observable 
is $\B$. A necessary and sufficient 
condition for perfect unambiguous discrimination of two probability 
distributions is their orthogonality in the usual scalar product 
of $\real^k$, 
$$
p\cdot q = \sum_j p_jq_j=0.
$$
where $j$ is a multiindex labeling the outcome sequences.
Only in this case the individual outcomes can be uniquely identified with 
a particular probability distribution. Our aim is to
investigate under which conditions on quantum observables the
unambiguous discrimination is possible, i.e.
for which observables there exists a state $\varrho$ such that
the resulting probabilities are mutually orthogonal.
The problem will be formalized in more details in the following  
sections.

\subsection{Equivalence of observables}\label{sec:equivalence}
 Let us think an observable as a box which has a row of leds indicating 
the possible measurement outcomes. When a measurement is performed, one of the leds is flashing. We assume that the leds do not have 
a particular specification, such as 'up' or 'down'. It is our choice
to attach some arbitrary but different labels to each of the leds 
before we perform measurements. From this point of view, the box would still be the 
same measurement apparatus if the labeling of the leds would have been 
chosen differently. For instance, we regard a Stern-Gerlach apparatus
pointing in $z$ direction to be the same as that one pointing into $-z$ direction,
since they differ only by different labelings of outcomes.

Based on this picture we can define the following
equivalence relation for observables. Let $\A$ and $\B$ be two observables with the outcome space 
$\Omega=\{\omega_1,\ldots,\omega_k\}$. We say that $\A$ and $\B$ 
are \emph{equivalent} if there is a permutation $\pi$ 
of the numbers $1,\ldots,k$ such that $\B_j=\A_{\pi(j)}$
for every $j=1,\ldots,k$. In other words, two observables $\A$ and $\B$ are equivalent if the collections of effects $\A_1,\dots,\A_k$ and $\B_1,\dots,\B_k$ are composed of same effects (with the same multiplicity). This relation is indeed an equivalence 
relation in the set of all observables with the outcome space $\Omega$.

\section{Perfect discrimination of observables}\label{sec:problem}
\subsection{Perfect discrimination problem}
Assume that we have an unidentified observable $\X$ which is known to be equivalent to some observable in the set $\cO=\{\A,\B,\C\ldots\}$ of (inequivalent) observables, all having the outcome space $\Omega=\{\omega_1,\ldots,\omega_k\}$. The task is to determine the correct equivalence class by performing measurements in some suitably chosen states. It is assumed that only measurements of $\X$ are allowed and one can do at most $n$ measurements. 

The formalization of this task is as follows. 
\begin{itemize}
\item[(a)] $n$ systems are prepared to a compound state $\varrho$, which is a state in the tensor product space $\hi^{\otimes n}$. We call $\varrho$ a \emph{probe state}.
\item[(b)] An $\X$-measurement is performed $n$ times, once for each system.  For each $\X$-measurement, we get a measurement outcome $\omega_j$. Therefore, a result of the $n$ measurements is an element 
$\omega_{\vj}\equiv(\omega_{j_1},\ldots,\omega_{j_n})$ 
from the product space $\Omega^n$. 
\item[(c)] The requirement of perfect discrimination 
is that from each measurement result 
$\omega_{\vj}$ occurring with 
non-zero probability, we can conclude the correct observable $\X$. This means that $\omega_{\vec{j}}$ can be obtained with non-zero probability only for a single observable in $\obs$. 
\end{itemize}

If the unknown observable $\X$ can be identified uniquely in $n$ repetitions,
we say that the set $\obs$ can be perfectly discriminated
in $n$ shots. The following example illustrates the perfect 
discrimination problem under consideration.

\begin{example}
Let $\hi=\complex^2$ (spin-1/2 system) and consider two measurements
$\sigma_x,\sigma_z$ corresponding to (ideal) Stern-Gerlach apparatuses
oriented in directions $x,z$ respectively. That is
\be
\nonumber
\sigma_z&\leftrightarrow&\A_\uparrow=|\uparrow_z\r\l \uparrow_z|,\quad 
\A_\downarrow=|\downarrow_z\r\l \downarrow_z|\, ; \\
\nonumber
\sigma_x&\leftrightarrow&\B_\uparrow=|\uparrow_x\r\l \uparrow_x|,\quad 
\B_\downarrow=|\downarrow_x\r\l \downarrow_x|\, ;
\ee 
where $|\uparrow_x\r=\frac{1}{\sqrt{2}}(|\uparrow_z\r +|\downarrow_z\r)$
and $|\downarrow_x\r=\frac{1}{\sqrt{2}}(|\uparrow_z\r -|\downarrow_z\r)$.
Because of the equivalence relation between the observables, the correct
labels are unknown and we cannot distinguish whether the shining led 
corresponds to 'up' or 'down'. If this would be possible, then we could distinguish between $\sigma_z$ and $-\sigma_z$. 
It follows that we must use the apparatus at least twice to 
find one of the following four pairs of outcomes: $\omega_{\uparrow\uparrow},
\omega_{\downarrow\downarrow}, \omega_{\uparrow\downarrow},
\omega_{\downarrow\uparrow}$. Morever, we are
able to distinguish only whether the outcomes are same
($\omega_{\uparrow\uparrow}$ or $\omega_{\downarrow\downarrow}$)
or different 
($\omega_{\uparrow\downarrow}$ or $\omega_{\downarrow\uparrow}$).
Consider a probe state
\be
\nonumber
|\psi\r&=&\frac{1}{\sqrt{2}} \left( |\uparrow_z\r\otimes|\uparrow_z\rangle
-|\downarrow_z\rangle\otimes|\downarrow_z\rangle \right) \\
\nonumber &=&
\frac{1}{\sqrt{2}} \left( |\uparrow_x\r\otimes|\downarrow_x\rangle
+|\downarrow_x\rangle\otimes|\uparrow_x\rangle \right) \, .
\ee
For $\sigma_x$ the probabilities of outcomes $\omega_{\uparrow\uparrow}$
and $\omega_{\downarrow\downarrow}$ vanishes. On the other hand, for $\sigma_z$
the outcomes $\omega_{\uparrow\downarrow}$ and $\omega_{\downarrow\uparrow}$
have zero probability. Thus, if we repeat the unknown measurement twice
and observe the same outcomes, it follows that $\X$ is $\sigma_z$. When the
outcomes are different, then $\X$ is $\sigma_x$.
\end{example}

In the general scheme described in (a)-(c), there are $k^n$ possible 
sequences of measurement results. However, as we distinguish only between 
different equivalence classes of observables, we have to group these 
sequences into subsets. Essentially, our conclusion are based only
on mutual relation of individual outcomes observed in $n$ repetitions
of the measurement. For instance, the measurement outcome sequences
\begin{equation}\label{eqn:same}
(\omega_1,\omega_1,\ldots,\omega_1),\ \ldots,\ (\omega_k,\omega_k,\ldots,\omega_k)
\end{equation}
must all lead to the same experimental result saying that all the outcomes 
are the same. Therefore, we cannot discriminate $k^n$ observables in $n$ 
shot measurement. To make this observation more explicit, 
let $S_k$ be the symmetric group of $k$ elements. For every 
$\pi\in S_k$ and $\omega_{\vec{j}}\equiv
\omega_{j_1,\dots,j_n}\equiv(\omega_{j_1},\ldots,\omega_{j_n})\in\Omega^n$, we denote
\begin{equation*}
\pi \cdot (\omega_{j_1},\ldots,\omega_{j_n}) 
= (\omega_{\pi(j_1)},\ldots,\omega_{\pi(j_n)})=\omega_{\pi(\vec{j})}.
\end{equation*}
This defines an action of the group $S_k$ on $\Omega^n$. The set $\Omega^n$ 
is thus composed as a disjoint union of the $S_k$-orbits. One of the orbits 
is clearly the set consisting of the sequences in (\ref{eqn:same}).

For a perfect discrimination of a set $\obs$ in $n$ shots, each orbit of $\Omega^n$ is either associated with a particular observable from $\cO$, or the probability for all elements in this orbit is zero. If $\obs$ consists of $M$ observables, then the set $\Omega^n$ is divided into $M$ disjoint 
subsets $\R_1,\dots,\R_{M}$ which are closed
under the action of the permutation group $S_k$. The subsets 
$\R_{1},\dots,\R_M$ indicate different discrimination results. (The orbits occuring with zero probability can be added to any subset $\R_j$ without affecting discrimination.) 
It follows that the maximal number of $k$-valued observables 
($k=|\Omega|$) that can be discriminated in $n$ shots is bounded 
by the the number of $S_k$-orbits in $\Omega^n$.

For example, for $n=3,k\ge 3$, the outcomes $\omega_{j_1j_2j_3}$ are grouped into five equivalence classes that can be denoted as $\omega_{xxx}$, $\omega_{xxy}$, $\omega_{xyx}$, $\omega_{xyy}$, $\omega_{xyz}$, where $x,y,z\in\Omega=\{1,2,\dots,k\}$ and $x\ne y\ne z\ne x$.
Thus, using three shots we can discriminate at most between 
five observables. The following example illustrates such discrimination
of five qutrit observables.

\begin{example}\label{ex:abcde}
Let $\hi=\complex^3$ and $\{\varphi_1,\varphi_2,\varphi_3\}$ be an orthonormal basis. Let $\A,\B,\C,\D,\E$ be the following four observables, each having three different outcomes:
\begin{eqnarray*}
\A_1 &=& \kb{\varphi_1}{\varphi_1},\quad \A_2 = \kb{\varphi_2}{\varphi_2},\quad \A_3 =\kb{\varphi_3}{\varphi_3};\\
\B_1 &=& \kb{\varphi_1}{\varphi_1},\quad \B_2 = \kb{\varphi_2}{\varphi_2}+\kb{\varphi_3}{\varphi_3},\quad \B_3 = O; \\
\C_1 &=& \kb{\varphi_1}{\varphi_1}+ \kb{\varphi_3}{\varphi_3},\quad \C_2 = \kb{\varphi_2}{\varphi_2},\quad \C_3 = O; \\
\D_1 &=& \kb{\varphi_1}{\varphi_1}+ \kb{\varphi_2}{\varphi_2},\quad \D_2 = \kb{\varphi_3}{\varphi_3},\quad \D_3 = O; \\
\E_1 &=& \kb{\varphi_1}{\varphi_1}+\kb{\varphi_2}{\varphi_2}+\kb{\varphi_3}{\varphi_3},\quad \E_2=\E_3=O.
\end{eqnarray*}
These observables can be discriminated with three measurements. Indeed, 
the probe state $\varrho$ corresponding to the vector 
$\varphi_1\otimes\varphi_2\otimes\varphi_3$ leads to the discrimination 
according to following table
\begin{center}
\begin{tabular}{|l|c|c|c|c|c|}
\hline  result & $xxx$ & $xxy$ & $xyx$ & $xyy$ & $xyz$ \\ 
\hline  conclusion & $\E$ & $\D$ & $\C$ & $\B$ & $\A$ \\
\hline
\end{tabular}
\end{center}
\end{example}

\subsection{Perfect discrimination of two observables with two shots}

Let us assume that an observable $\X$ 
is known to be equivalent to either $\A$ or $\B$. 
For each state $\varrho$, we denote by 
$p^\A_\varrho(\omega_{\vec{j}})$ and 
$p^\B_\varrho(\omega_{\vec{j}})$ the probabilities of 
getting the outcomes $\omega_{\vec{j}}=(\omega_{j_1},\ldots,\omega_{j_n})$ 
in $n$ repetitions of $\A$-measurements or $\B$-measurements, respectively. 

As already discussed in Section \ref{sec:iden_obs},  the problem of perfect discrimination of observables is reduced to a perfect discrimination of the corresponding probability 
distributions. In order to discriminate $\A$ and $\B$ in $n$-shots, we need to find a probe state $\varrho$ such that
\begin{equation}\label{eqn:basic}
p^\A_\varrho(\omega_{\vec{j}})
\ p^\B_\varrho(\omega_{\pi(\vec{j})})=0 
\quad\forall\omega_{\vec{j}}\in\Omega^n,\quad\forall\pi\in S_k\, .
\end{equation}

The probabilities under consideration have the form
\begin{equation}
p^\A_\varrho(\omega_{j_1},\ldots,\omega_{j_n})=\tr{\varrho \A_{j_1}\otimes\cdots\otimes\A_{j_n}}.
\end{equation}
Hence, if 
\begin{equation}
p^\A_\varrho(\omega_{j_1},\ldots,\omega_{j_n})=0
\end{equation}
for some state $\varrho$, then there is a unit vector $\psi\in\hi$ such that 
\begin{equation}\label{eqn:psi}
\ip{\psi}{\A_{j_1}\otimes\cdots\otimes\A_{j_n}\psi}=0 \, .
\end{equation}
This is easily seen by decomposing $\varrho$ to a convex combination of pure states. We thus conclude that the search for suitable probe state $\varrho$ can be restricted to pure states. Moreover, since the operator $\A_{j_1}\otimes\cdots\otimes\A_{j_n}$ is positive, $\psi$ in equation (\ref{eqn:psi}) must be its eigenvector with eigenvalue 0.

Let us consider a situation where we have only two systems available and hence, we are trying to discriminate observables by performing two measurements. The set $\Omega^2$ has two orbits under the action of $S_k$. One orbit consists of the following pairs: $(\omega_1,\omega_1),\ (\omega_2,\omega_2), \ \ldots,\ (\omega_k,\omega_k)$. The other orbit contains all the other pairs, i.e., $(\omega_i,\omega_j)$ with $i\neq j$. As there are only two orbits, we cannot discriminate more than two observables.

In order to discriminate two observables $\A$ and $\B$ with two shots there must be a state $\varrho$ such that either
\begin{eqnarray}
p^\A_\varrho(\omega_i,\omega_i) &=& 0 \quad \forall i=1,\ldots,k \label{eqn:A0}\\
p^\B_\varrho(\omega_i,\omega_j) &=& 0 \quad \forall i,j=1,\ldots, k,\  i\neq j \label{eqn:B0}
\end{eqnarray}
or the same conditions with $\A$ and $\B$ interchanged.

In terms of operators the above conditions read
\begin{eqnarray}
\tr{\varrho \A_j\otimes\A_j} &=& 0 \quad \forall i=1,\ldots,k \label{eqn:A0op}\\
\tr{\varrho \B_i\otimes\B_j} &=& 0 \quad \forall i,j=1,\ldots, k,\  i\neq j \label{eqn:B0op}
\end{eqnarray}
Still another set of equivalent conditions is the following:
\begin{eqnarray}
\sum_{j=1}^k \tr{\varrho \A_j\otimes\A_j} &=& 0 \label{eqn:sumA0} \\
\sum_{j=1}^k \tr{\varrho \B_j\otimes\B_j} &=& 1. \label{eqn:sumB1} 
\end{eqnarray}
The last equation follows from the normalization condition
$\sum_{i,j} \B_i\otimes\B_j= \sum_i \B_i  \otimes \sum_j \B_j = I$.

\begin{proposition}\label{prop:zeros}
If $\A$ and $\B$ can be discriminated with two shots, then the operators $\A_1,\ldots,\A_k,\B_1,\ldots,\B_k$, except possibly one of them, have eigenvalue 0.
\end{proposition}

\begin{proof}
Assume that conditions (\ref{eqn:A0op}) and (\ref{eqn:B0op}) hold, the other case (i.e. $\A$ and $\B$ interchanged) being similar.  Since all the tensor product operators $\A_1\otimes\A_1,\ldots,\A_k\otimes\A_k$ have eigenvalue zero, it follows that all the operators $\A_1,\A_2,\ldots,\A_k$ have eigenvalue 0. Assume now that, for instance, $\B_1$ does not have eigenvalue 0. Since all the tensor product operators $\B_1\otimes\B_2,\ldots,\B_1\otimes\B_k$ have eigenvalue zero, this means that the operators $\B_2,\ldots,\B_k$ have eigenvalue 0.
\end{proof}

In the next example we demonstrate that the exception mentioned in Proposition \ref{prop:zeros} is indeed possible: one of the operators $\A_1,\ldots,\A_k,\B_1,\ldots,\B_k$ need not to have eigenvalue 0.

\begin{example}\label{ex:noname}
Let $\hi=\complex^3$, $\{\varphi_1,\varphi_2,\varphi_3\}$ an orthonormal basis and $0<t<1$. We define two observables $\A$ and $\B$ with the outcome space $\{1,2\}$ as
\begin{eqnarray*}
\A_1&=&\kb{\varphi_1}{\varphi_1},\\
\A_2 &=& \kb{\varphi_2}{\varphi_2}+\kb{\varphi_3}{\varphi_3},\\
\B_1&=&\kb{\varphi_1}{\varphi_1}+\kb{\varphi_2}{\varphi_2}+t\kb{\varphi_3}{\varphi_3}, \\
\B_2 &=& (1-t)\kb{\varphi_3}{\varphi_3}.
\end{eqnarray*}
The operator $\B_1$ has eigenvalues $t$ and 1. The observables $\A$ and $\B$ can be clearly discriminated with the state corresponding to the vector $\varphi_1\otimes\varphi_2$. 
\end{example}

\subsection{Perfect discrimination of two qubit observables with two shots}
\label{sec:qubit22}
Let $\A$ and $\B$ two observables defined on $\complex^2$ and having $k$ possible outcomes $\{\omega_1,\ldots,\omega_k\}$. In the following we derive a necessary and sufficient condition for $\A$ and $\B$ to be perfectly discriminable. 

First of all, if an operator on $\complex^2$ has eigenvalue 0, then it is a multiple of a one dimensional projection. Thus, Proposition \ref{prop:zeros} implies that
\be
\nonumber \A_j&=&a_j |\varphi_j\r\l\varphi_j|  \ \ \ \forall\, j\\
\nonumber \B_j&=& b_j |\phi_j\r\l\phi_j| \ \ \ \forall\, j<k\\
\nonumber \B_k&=&I-\sum_j \B_j\, ,
\ee
where $\varphi_j,\phi_j$ are unit vectors and $a_j,b_j\in [0,1]$. 
It then follows from (\ref{eqn:A0op}) and (\ref{eqn:B0op}) that a pure probe state
$\varrho=|\psi\r\l\psi|$ must be orthogonal to
states $\varphi_j\otimes\varphi_j$ and 
$\phi_i\otimes\phi_j$ for all $i\ne j$, provided that $a_j\ne 0$ and $b_i\ne 0 \ne b_j$. Whenever there are at least three different states $\varphi_j$, then the
vectors $\varphi_j\otimes\varphi_j$ are linearly independent and span a three dimensional subspace. The one dimensional subspace orthogonal to that subspace
is spanned by the singlet state. 
However, the singlet state has nonzero overlap with $\phi_i\otimes\phi_j$ if $\phi_i$ and $\phi_j$ are not collinear. Unless the measurement $\B$ is trivial, there is at least one pair of such vectors.

We conclude that a necessary criterion for the discrimination of $\A$ and $\B$ is that there are only two different states $\varphi_1,\varphi_2$ in the range of $\A$. Thus,
each operator $\A_j$ is proportional either to $\kb{\varphi_1}{\varphi_1}$ or $\kb{\varphi_2}{\varphi_2}$. Moreover, the normalization $\sum_j \A_j=I$ requires that these states are orthogonal so we denote them $\varphi \equiv \varphi_1$ and $\varphi_\perp \equiv \varphi_2$. Consequently, the probe state $\varrho=|\psi\r\l\psi|$
must be of the form
\be
\psi=\alpha\varphi\otimes\varphi_\perp+\beta\varphi_\perp\otimes\varphi
\ee
for some $\alpha,\beta\in\complex,\mod{\alpha}^2+\mod{\beta}^2=1$.

Let us now fix two indices $i$ and $j$ such that $i\neq j$ and $b_i\ne 0\ne b_j$
. The orthogonality relations $0=\l\phi_i\otimes\phi_j|\psi\r=
\l\phi_j\otimes\phi_i|\psi\r$ can then be written in the form  
\be
0&=&\alpha\l\phi_i|\varphi\r\l\phi_j|\varphi_\perp\r+\beta\l\phi_i|\varphi_\perp\r\l\phi_j|\varphi\r\\
0&=&\beta\l\phi_i|\varphi\r\l\phi_j|\varphi_\perp\r+\alpha\l\phi_i|\varphi_\perp\r\l\phi_j|\varphi\r\, .
\ee
By expressing the states $\phi_i,\phi_j$ in the basis $\varphi,\varphi_\perp$ we get
\be
\phi_i&=& a\varphi+e^{ir}\sqrt{1-a^2}\varphi_\perp\, \\
\phi_j&=& b\varphi+e^{is}\sqrt{1-b^2}\varphi_\perp\, .
\ee
Here $a,b\in [0,1]$ and $r,s\in [0,2\pi)$.
In this notation we have
\be
0=\alpha a\sqrt{1-b^2}e^{-is}+\beta b\sqrt{1-a^2} e^{-ir}\, ,\\
0=\beta a\sqrt{1-b^2}e^{-is}+\alpha b\sqrt{1-a^2} e^{-ir}\, ,
\ee
and consequently
\be
(a\sqrt{1-b^2}e^{-is}+b\sqrt{1-a^2} e^{-ir})(\alpha+\beta)=0\, ,\\
(a\sqrt{1-b^2}e^{-is}-b\sqrt{1-a^2} e^{-ir})(\alpha-\beta)=0\, .
\ee

There are two possible solutions: 
\be
{\rm either}\ \ a=b\, , & \alpha=-\beta=1/\sqrt{2}\, , & r=s \, ; \label{sol:first}\\ 
{\rm or}\ \ a=b\, , & \alpha=\beta=1/\sqrt{2}\, ,  & r=s+\pi \, . \label{sol:second}
\ee
Here we use the facts that $\mod{\alpha}^2+\mod{\beta}^2=1$ and that $\alpha$ and $\beta$ can be multiplied by a common phase factor without changing the probe state $\varrho$.

The first solution (\ref{sol:first}) leads to the trivial situation where for all $j$,
\be
\phi_j\equiv\phi=a\varphi+e^{ir}\sqrt{1-a^2}\varphi_\perp\, .
\ee
Then the normalization condition 
$\sum_j\B_j=\sum_j b_j |\phi\r\l\phi|\ne I$ does not hold and therefore 
this case does not represent a valid solution.

The second solution (\ref{sol:second}) implies that for every pair $i\neq j$ (with $b_i\ne 0\ne b_j$), we get
\be
\phi_i=a\varphi+e^{ir}\sqrt{1-a^2}\varphi_\perp \equiv \phi_+\, ,\\ 
\phi_j=a\varphi-e^{ir}\sqrt{1-a^2}\varphi_\perp \equiv \phi_-\, .
\ee
In particular, this means that there can be only two nonzero $b_j$. Without lost of generality, we assume that $b_1\neq 0 \neq b_2$ and $b_3=\ldots =b_k=0$, i.e.,  $\B_1=b_1\kb{\phi_+}{\phi_+}$, $\B_2=b_2\kb{\phi_-}{\phi_-}$, $\B_3=\ldots=\B_k=O$.
In order to fulfill 
the normalization constraint $\B_1 + \B_2=I$ they must
correspond to mutually orthogonal projectors, i.e. $b_1=b_2=1$ and
$\l\phi_+|\phi_-\r=0$. Since $\l\phi_+|\phi_-\r=2a^2-1$, we get
\be
\phi_\pm=\frac{1}{\sqrt{2}}(\varphi\pm e^{ir}\varphi_\perp)\, . 
\ee
We summarize this solution in the following proposition.

\begin{proposition}\label{prop:3}
Two qubit observables $\A$ and $\B$ can be perfectly discriminated in two shots
only if they are of the following form
\begin{equation*}
\begin{array}{ccc}
\A_1=a_1|\varphi\r\l\varphi| & & \B_1=|\phi_+\r\l\phi_+| \\
\vdots & & \B_2=|\phi_-\r\l\phi_-|  \\
\A_{m}=a_{m}|\varphi\r\l\varphi|  & & \B_3=O \\
\A_{m+1}=a_{m+1}|\varphi_\perp\r\l\varphi_\perp| 
& & \vdots\\
\vdots & & \vdots \\
\A_k=a_k|\varphi_\perp\r\l\varphi_\perp|  & & \B_k=O
\end{array}
\end{equation*}
with $\sum_{j\le m}a_j=\sum_{j> m}a_j=1$,
and
\begin{equation*}
\phi_\pm=\frac{1}{\sqrt{2}}(\varphi\pm e^{ir}\varphi_\perp)\, 
\end{equation*}
for some $r\in [0,2\pi)$.
The probe state is
$$
\psi=\frac{1}{\sqrt{2}}(\varphi\otimes\varphi_\perp
+\varphi_\perp\otimes\varphi).
$$
\end{proposition}

\section{Unambiguous discrimination and identification of observables}
\label{sec:unambiguous}
\subsection{Unambiguous identification problems}
In the previous sections we have investigated perfect discrimination of quantum observables. We have seen that this is possible only in some special cases.
Therefore, generalizations of the perfect discrimination scheme are needed.

In what follows we will generalize the framework of perfect discrimination and introduce the concepts of unambiguous discrimination and unambiguous identification of quantum observables. In these generalizations all conclusions are still required to be error-free, but also inconclusive results are allowed. Moreover, it is not assumed that there is a conclusive result for each a priori possibility. 
As before, the starting point is that we are given a measurement apparatus $\X$ which is known to be described by an observable from the set $\obs=\{\A,\B,\C,\ldots\}$. The goal
is to single out the correct observable. We are interested in the following four situations:  
\begin{itemize}
\item[(PD)]  Perfect discrimination of the set $\obs$ means that we can deduce $\X$ from any measurement result occurring with nonzero probability. There are no inconclusive results.
\item[(UD)] Unambiguous discrimination of the set $\obs$ means that whichever $\X$ is, we have a nonzero probability to arrive to a conclusion.
\item[(PI)] Perfect identification of a subset $\obs'\subset\obs$ from $\obs$ means that whatever measurement result we get, we can conclude whether $\X$ is $\A$ or not for each $\A\in\obs'$.
\item[(UI)] Unambiguous identification of a subset $\obs'\subset\obs$ from $\obs$ means that if $\X=\A\in\obs'$, then there is a nonzero probability to get a measurement result leading to this conclusion.
\end{itemize}
It is clear that UI is the most general scheme of these and the other three are special cases of it. PI becomes PD and UI becomes UD when $\obs'=\obs$. UD reduces to PD when the probability of inconclusive result is zero.

Let us consider a situation where we make $n$ measurements with the unknown measurement apparatus $\X$. The total outcome space is thus $\Omega^n$, and we divide it into disjoint subsets $\R_1,\R_2,\ldots,$ and $\R_?$. The last subset $\R_?$ is associated with the inconclusive result; if an outcome from $\R_?$ is recorded, we cannot make a conclusion. The other subsets $\R_1,\R_2,\ldots,$ correspond to conclusions $\X=\A,\X=\B$ etc. as in the case of perfect discrimination. A probability for each conclusive outcome must vanish for all observables except one of them. If, for instance, the subset $\R_1$ is associated with the conclusion $\X=\A$, then for every $\omega_{\vj}\in\R_1$, we must have
\be
p^\A_\varrho(\omega_{\vec{j}})\neq 0=p^\B_\varrho(\omega_{\vec{j}})=p^\C_\varrho(\omega_{\vec{j}})=\ldots.
\ee

\begin{proposition}\label{prop:no-unambi-disc}
If an observable $\A$  can be unambiguously discriminated, then at least one effect $\A_j$ has eigenvalue zero.
\end{proposition}

\begin{proof}
In order to have zero probability for a result $\omega_{\vj}=(\omega_{j_1},\dots,\omega_{j_n})$, the corresponding
effect $\A_{\vec{j}}=\A_{j_1}\otimes\dots\otimes\A_{j_n}$ must have
at least one zero eigenvalue. The eigenvalues of $\A_{\vec{j}}$ are products of the eigenvalues of the effects $\A_j$, hence at least one of the effects 
$\A_j$ must have eigenvalue zero. 
\end{proof}
Let us note that the above impossibility statement holds only for 
unambiguous discrimination problem and it is not applicable to
general unambiguous identification problem. In fact, although $\A$
cannot be unambiguously discriminated, it 
can still be unambiguously identified as we will see in 
the end of Subsection \ref{sec:4.2}.

For an inconclusive result (i.e. an outcome belonging to $\R_?$) 
the probability is nonzero for more than one observable. 
The goal of a given task for a given set of observables $\obs$ is to decide
on the existence of a suitable probe state $\varrho$ 
and maximize (over all potential probe states) the average probability 
of getting the conclusive results (success probability)
$P_{\rm succ}^\varrho=\sum_{\X\in\obs}\eta^\X p_{\varrho}^\X(\R_\X)$, where
$\eta:\obs\to[0,1]$ is a given probability distribution of the elements
in $\obs$ reflecting our apriori information, i.e. $\sum_{\X\in\obs}\eta^\X=1$.

\subsection{Unambiguous identification of sharp qubit observables with two shots}
\label{sec:4.2}

By making two measurements with the unknown measurement apparatus $\X$ we get only two results; either the outcomes are same or different.
Therefore, we have three options: 
\begin{itemize}
\item both results are conclusive (perfect discrimination of two observables);
\item one of the results is conclusive and the second one is inconclusive (unambiguous identification of one observable);
\item both results are inconclusive (no identification at all).
\end{itemize}

Let us assume that $\X$ is known to be either $\A$ or $\B$, which are both sharp qubit observables. Proposition \ref{prop:3} implies that perfect discrimination of $\A$ and $\B$ is possible if and only if there are of the form
\be
\begin{array}{rclcrcl}
\A_1 &=& |\varphi\r\l\varphi|\, , &\quad& \A_2 &=& |\varphi_\perp\r\l\varphi_\perp| \, , \\
\B_1 &=& |\phi_+\r\l\phi_+|\, , &\quad& \B_2 &=& |\phi_-\r\l\phi_-| \, ,
\end{array}
\ee
where $\phi_\pm=\frac{1}{\sqrt{2}}(\varphi\pm e^{ir}\varphi_\perp)$ for some $r\in\real$.

In the following we investigate unambiguous identification of two sharp qubit observables in two shots. There are only two results in two shot measurement scheme: $\Rsame$ and $\Rdiff$. We have thus two choices; either $\Rsame$ is conclusive or $\Rdiff$ is conclusive. For sharp qubit observables these two options are equally good as far as we consider the success probability. Indeed, this observation is proved below in Proposition \ref{prop:same-diff}.   

We recall that a sharp qubit observable $\A$ is described, up to equivalence, by a unit vector $\va\in\real^3$. Namely, a sharp qubit observable $\A$ with two outcomes $\omega_1,\omega_2$ is given by
\be\label{eq:simplequbitA}
\begin{array}{rcl}
\omega_1\mapsto \A_1&=& \half \left( \id + \va\cdot\vsigma \right) \\
 \omega_2\mapsto \A_2 &=& \half \left( \id - \va\cdot\vsigma \right).
\end{array}
\ee 

\begin{proposition}\label{prop:same-diff}
Let $\A$ and $\B$ be sharp qubit observables
apriori distributed according to
probabilities $\eta^\A=\eta$ and $\eta^\B=1-\eta$.
The following statements are equivalent:
\begin{itemize}
\item[(i)] There is a probe state $\psi$ such that $\Rsame$ leads to the conclusion $\X=\A$ and the success probability is $P_{\rm succ}$.
\item[(ii)] There is a probe state $\psi'$ such that $\Rdiff$ leads to the conclusion $\X=\A$ with the success probability $P_{\rm succ}$.
\end{itemize}
\end{proposition}

\begin{proof}
First statement asserts that 
\be
\nonumber
P_{\rm succ}&=& \eta \l\psi|\A_1\otimes\A_1+\A_2\otimes\A_2|\psi\r \, ,\\
\nonumber
0&=&\l\psi|\B_1\otimes\B_1+\B_2\otimes\B_2|\psi\r\, .
\ee
Let $\A_1=\frac{1}{2}(I+\va\cdot\vsigma)$ 
and $\B_1=\frac{1}{2}(I+\vb\cdot\vsigma)$. The selfadjoint unitary operator 
 $U:=\frac{\va\times\vb}{\no{\va\times\vb}}\cdot\vsigma$ transforms $\A_1$ and $\B_1$ in the following way:
 $U\A_1U=\A_2$, $U\B_1U=\B_2$. Thus, defining $\psi^\prime=(I\otimes U)\psi$ we obtain
\begin{eqnarray*}
P_{\rm succ} &=& \eta \l\psi|\A_1\otimes\A_1+\A_2\otimes\A_2|\psi\r \\
&=& \eta \l\psi^\prime|\A_1\otimes\A_2+\A_2\otimes\A_1|\psi^\prime\r 
\end{eqnarray*}
and
\begin{eqnarray*}
0 &=& \l\psi|\B_1\otimes\B_1+\B_2\otimes\B_2|\psi\r \\
&=& \l\psi^\prime|\B_1\otimes\B_2+\B_2\otimes\B_1|\psi^\prime\r\, .
\end{eqnarray*}
\end{proof}

Let $\A$ and $\B$ be two sharp qubit observables. 
Assume that $\R_{\rm diff}$ is the inconclusive result 
and $\R_{\rm same}$ is the conclusive result $\X=\A$. This means that
\be
\tr{ \varrho\A_j\otimes \A_j}\neq 0 &\quad&  \textrm{for $j=1$ or $j=2$}  \label{un_dis_1} \\
\tr{\varrho\B_j\otimes \B_j}= 0 &\quad& \textrm{for $j=1,2$} .
\ee
Let $\phi$ and $\phi_\perp$ be orthogonal unit vectors such that $\B_1=\kb{\phi}{\phi}$ and $\B_2=\kb{\phi_\perp}{\phi_\perp}$. From the second condition it follows that a probe state $\psi$ is of the form
\be
\psi=\alpha \phi\otimes\phi_\perp + \beta \phi_\perp\otimes\phi \, 
\ee
for some $\alpha,\beta\in\complex$, $|\alpha|^2+|\beta|^2=1$.
Inserting this state into (\ref{un_dis_1}) we obtain the following expression
for the probability $P_{\rm succ}$ of conclusive result:
\be
P_{\rm succ}&=& \eta \l \psi|\A_1\otimes\A_1+\A_2\otimes\A_2|\psi\r\\
&=& \frac{1}{2} \eta \left( 1+\l\psi|\va\cdot\vsigma
\otimes\va\cdot\vsigma|\psi\r \right) \, ,
\ee
where we have denoted $\A_1=\frac{1}{2}(\id+\va\cdot\vsigma)$. In order
to maximize the above probability one needs to maximize the
term 
\begin{eqnarray*}
\l\psi|\va\cdot\vsigma\otimes\va\cdot\vsigma
|\psi\r &=& (\alpha^*\beta+\beta^*\alpha)|\l\phi|\va\cdot\vsigma|\phi_\perp\r|^2 \\
& & -\ip{\phi}{\va\cdot\vsigma|\phi}^2\, .
\end{eqnarray*}
This expression achieves maximum for $\alpha=\beta=1/\sqrt{2}$, thus the
optimal probe state is
\be
\psi=\frac{1}{\sqrt{2}}(\phi\otimes\phi_\perp + \phi_\perp\otimes\phi)\, 
\ee
and
\begin{equation*}
P_{\rm succ}=\frac{1}{4}\eta [3+\no{\va\times\vb}^2-3(\va\cdot\vb)^2]=\eta \sin^2\theta_{\va\vb},
\end{equation*}
where $\theta_{\va\vb}$ is the angle between $\va$ and $\vb$.

\begin{proposition}
If $\A$ and $\B$ are sharp qubit observables, the success probability $P_{\rm succ}$ of 
unambiguous identification of $\A$ in two shots is
\be
P_{\rm succ}=\eta \sin^2\theta_{\va\vb}
\, .
\ee
\end{proposition}

This result can be generalized to a pair of a sharp observable $\B$ 
and an unsharp observable $\A$ defined by a vector $\va\in\real^3$ 
($\no{\va}\leq 1$) through formula (\ref{eq:simplequbitA}).
In such case $\A$ can be identified unambiguously and
Proposition~\ref{prop:same-diff} holds as $\A_1$ and $\A_2$ are connected by a unitary transformation. The optimal probe state is the same as previously and for the success probability we get
\be
P_{\rm succ}=\eta \left( \no{\va}^2 \sin^2\theta_{\va\vb}+\half (1-\no{\va}^2) \right).
\ee
In particular, $P_{\rm succ}\neq 0$ whenever $\A$ and $\B$ are inequivalent. Note that if $\no{\va}<1$, then the operators $\A_1$ and $\A_2$ do not have eigenvalue 0. Even though Proposition \ref{prop:no-unambi-disc} implies that $\A$ cannot be discriminated, we have seen that it can be identified.

\subsection{Unambiguous discrimination of sharp qubit observables}

As explained in the beginning of the previous section, unambiguous discrimination in two shots is possible only in the case of perfect discrimination of two observables. Hence, for two sharp qubit observables $\A$ and $\B$ which do not satisfy condition $\va\perp\vb$, we need at least three shots for their unambiguous discrimination.

One possibility is to perform twice the procedure of unambiguous identification in two shots, once to identify $\A$ and once for $\B$. If we do this in the optimal way as characterized in Subsection \ref{sec:4.2}, then we get the following result.

\begin{proposition}
Unambiguous discrimination of two sharp qubit observables $\A$ and $\B$ is possible in four shots and
\be\label{eqn:four}
P_{\rm succ} \geq \sin^2\theta_{\va\vb}
\, .
\ee
\end{proposition}

In particular, the probe state is 
\be
\psi=\frac{1}{2}(\phi\otimes\phi_\perp + \phi_\perp \otimes\phi)
\otimes
(\varphi\otimes\varphi_\perp+\varphi_\perp\otimes\varphi)
\ee
where $\{\varphi,\varphi_\perp\}$, $\{\phi,\phi_\perp\}$ are bases associated
with sharp observables $\A,\B$, respectively. First pair of outcomes allows
us to unambiguously identify the observable $\A$ and the second pair 
of outcomes unambiguously identifies the observable $\B$. Both conclusions
happen with probability $\sin^2\theta_{\va\vb}$. Therefore the average success
probability achieves just the same value whatever is the initial
distribution of observables $\A$ and $\B$. We leave it as an open problem 
whether the equality holds in (\ref{eqn:four}) and also whether unambiguous 
discrimination of $\A$ and $\B$ is possible in three shots.

\section{Conclusions}\label{sec:conclusions}
In this paper we developed the general framework, in which
different variations of unambiguous identification tasks
for quantum observables can be tackled. In all the problems considered here, we are given an unknown apparatus $\X$ promised to be
one from a given finite set of observables. The goal is to identify the
observable without an error. Moreover, we are interested in minimal
resources necessary for the successful realization while keeping the
success probability as large as possible. Resources are quantified
in a number of probe systems, i.e. usages of the unknown apparatus. Nothing can be concluded if the apparatus is used only once, hence the minimal number of usages
is two. Because of the unknown labeling of the given measurement device 
the discrimination cannot be based on particular 
outcome sequence, but rather on its symmetry. For instance, in two 
shots scenario we can only say whether the outcomes are different
or same.

We formulated the problems in general settings and presented some solutions in the case qubit observables. We succeeded to show that
using the unknown measurement device twice we can perfectly discriminate
only sharp qubit observables $\A,\B$ corresponding to Stern-Gerlach apparatuses
oriented in mutually orthogonal directions $\va\perp\vb$. For general
pair of sharp qubit observables only unambiguous conclusions are possible with two shots. In particular, using the apparatus twice we can conclusively 
identify only one of the observables (say $\A$) with probability 
$P_{\rm succ}=\eta\sin^2\theta_{\va\vb}$, where 
$\eta$ is a priori probability of $\A$ and
$\theta_{\va\vb}$ is the angle between directions $\va$ and $\vb$.
The value of $\sin^2\theta_{\va\vb}$ serves
also as the lower bound for the success probability of
the unambiguous discrimination of $\A$ and 
$\B$, in which both observables are identified conclusively. 
Interestingly, in all these cases the optimal probe state is a specific 
maximally entangled state. However, as shown in Examples \ref{ex:abcde} and \ref{ex:noname}, entangled states are not always necessary and also factorized states can be exploited for perfect discrimination.

Discrimination and identification type of problems are of interest,
because in these situations also the individual outcomes can provide 
us with useful information about unknown quantum apparatuses. 
This paper represents a preliminary step towards understanding of mutual
experimental distinguishability of quantum observables and 
callibration of quantum measurement devices. 
There are many interesting questions in this subject
deserving further investigation. 

\section*{Acknowledgements}
This work was supported by the European Union projects QAP,
CONQUEST, by the Slovak Academy of Sciences via the project CE-PI,
and by the projects APVV and VEGA. Authors 
wish to thank Vlado Bu\v{z}ek for inspiring discussions.


\end{document}